# Digital Twin Technologies in Predictive Maintenance: Enabling Transferability via Sim-to-Real and Real-to-Sim Transfer

Sizhe Ma,[1] Katherine A. Flanigan, Ph.D.,[2] and Mario Bergés, Ph.D.[3]

[1]Department of Civil and Environmental Engineering, Carnegie Mellon University, 5000 Forbes Ave, Pittsburgh, PA 15213; e-mail: sizhem@andrew.cmu.edu
[2]Department of Civil and Environmental Engineering, Carnegie Mellon University, 5000 Forbes Ave, Pittsburgh, PA 15213; e-mail: kflaniga@andrew.cmu.edu (Corresponding author)
[3]Department of Civil and Environmental Engineering, Carnegie Mellon University, 5000 Forbes Ave, Pittsburgh, PA 15213; e-mail: mberges@andrew.cmu.edu

Mario Bergés holds concurrent appointments at Carnegie Mellon University (CMU) and as an Amazon Scholar. This manuscript describes work at CMU and is not associated with Amazon.

## ABSTRACT

The advancement of the Internet of Things (IoT) and Artificial Intelligence has catalyzed the evolution of Digital Twins (DTs) from conceptual ideas to more implementable realities. Yet, transitioning from academia to industry is complex due to the absence of standardized frameworks. This paper builds upon the authors' previously established functional and informational requirements supporting standardized DT development, focusing on a crucial aspect: transferability. While existing DT research primarily centers on asset transfer, the significance of "sim-to-real transfer" and "real-to-sim transfer"—transferring knowledge between simulations and real-world operations— is vital for comprehensive lifecycle management in DTs. A key challenge in this process is calibrating the "reality gap," the discrepancy between simulated predictions and actual outcomes. Our research investigates the impact of integrating a single Reality Gap Analysis (RGA) module into an existing DT framework to effectively manage both sim-to-real and real-to-sim transfers. This integration is facilitated by data pipelines that connect the RGA module with the existing components of the DT framework, including the historical repository and the simulation model. A case study on a pedestrian bridge at Carnegie Mellon University showcases the performance of different levels of integration of our approach with an existing framework. With full implementation of an RGA module and a complete data pipeline, our approach is capable of bidirectional knowledge transfer between simulations and real-world operations without compromising efficiency.

## INTRODUCTION

The advancement of the Internet of Things (IoT) and Artificial Intelligence (AI) technologies has rapidly transformed the concept of the Digital Twin (DT) from a visionary idea into a tangible reality. DTs are virtual models mirroring the structure, context, and behavior of physical systems, regularly updated with real-world data to outcomes and guide value-driven decisions. They are deeply rooted in the bidirectional interaction between the virtual and the physical (National Academy 2023). These virtual constructs are often realized through sets of models, offering capabilities extending beyond traditional modeling approaches that typically rely on limited models or sources of information. These capabilities include accurate replication, enhanced simulation, and advanced visualization, collectively providing multidimensional representations of the physical entity in virtual space. Given the field of civil engineering's comprehensive nature, with various applications such as structural health monitoring and infrastructure management, there is growing research exploring how DTs can revolutionize task execution and decision-making processes in the discipline.

However, the journey from academic exploration to industry adoption remains challenging, due to the lack of standardized frameworks across industries. Each application has adapted DT technologies to meet its unique requirements, leading to a fragmented landscape where a universally accepted definition—let alone a standardized computational framework—remains elusive. There is a need to accurately identify and define the underlying requirements supporting DTs, which guides the development of a more universal DT framework. Such a framework ensures all stakeholders' diverse needs and expectations are adequately addressed.

Ma et al. 2023 set the foundation for such a roadmap, focusing on DTs through the lens of asset maintenance, which is critical and universally relevant across civil engineering. This is not only due to its widespread application but also because it exemplifies the necessity for precision, efficiency, and collaborative decision-making, as well as the transition from reactive to proactive approaches—namely Predictive Maintenance (PMx)—that DTs are uniquely positioned to enhance (Ma et al. 2023). In Ma et al. 2023, a state-of-the-art review of PMx is conducted to identify and define the necessary Informational Requirements (IRs) and Functional Requirements (FRs) supporting DT automation. These requirements act as the established foundation for all stakeholders in the PMx pipeline, facilitating a clear understanding of each party's responsibilities and the appropriate channels for seeking specific system information or feature implementation. This shared framework ensures that stakeholders are aligned in their roles, responsibilities, and expectations, streamlining the collaborative process and enhancing the overall efficiency and effectiveness of the PMx system (Flanigan et al. 2022). To be universally accepted and effective across stakeholders, PMx DTs must fulfill each of the identified IRs and FRs. Ma et al. 2023's review defines 14 foundational IRs and FRs, including critical elements such as integration of physical properties, interpretability, and robustness. Due to the sheer volume of requirements, it is recommended to progressively tackle these challenges.

Transferability, a vital but lacking requirement, refers to the capability of a PMx system to adapt its predictive functions across different assets or varying conditions. This feature is crucial as many industries plan to use DTs for fleet management. However, transferability should not be limited to just asset transfer. As DTs were initially envisioned as holistic solutions guiding an asset's entire lifecycle, the transfer of knowledge between different lifecycle phases, such as from design to operation, is equally crucial (Grieves and Vickers 2017). Transferring knowledge between different lifecycle phases involves two types of challenges: "sim-to-real transfer" and "real-to-sim transfer" (Müller et al. 2022). The difficulty in transferring knowledge between domains lies in the "reality gap," the divergence between simulated and real-world environments. DTs are not merely tools for analysis and prediction; they are detailed virtual models of physical systems, created to reflect and engage with their real-world equivalents in real time. Minor differences in each component of the simulated model compared to actual conditions can accumulate, leading to substantial variations in outcomes (Stocco et al. 2023). The lack of effective methods to identify, measure, and manage the reality gap remains a significant hurdle impeding transferability.

This study investigates the impact of integrating the reality gap analysis (RGA) module within the existing DT framework. Our aim is to determine whether this integration enables knowledge transfer between simulation and real-world operations. To contextualize our analysis, we integrate the proposed work into the DT framework outlined by Gratius et al. 2024 (Gratius et al. 2024). To address the identified reality gap challenges in fulfilling knowledge transfer between two domains, our approach is characterized by three key aspects. First, we incorporate a confidence-based approach to efficiently quantify the reality gap between simulation data and real-world data for individual assets. Second, the quantified reality gaps are used to adjust the simulation data to better reflect individual assets in the real world. This step enables sim-to-real transfer. Third, we broaden our historical repository by reverse-applying the quantified reality gap to real-world data collected under critical conditions. This process neutralizes the impact of real-world discrepancies, enriching the repository with a more comprehensive knowledge base for the simulation domain, thus enabling real-to-sim transfer. We implement this approach in a condition-based monitoring task on the Newell-Simon Bridge, a short-span steel pedestrian truss bridge at Carnegie Mellon University. The integration of our approach and the existing framework is divided into three levels, referred to as the levels of integration (LoI), which span from Gratius et al. 2024's framework to the full implementation of the RGA module with the complete data pipeline. The comparison between LoIs indicates that our approach is capable of both sim-to-real transfer and real-to-sim transfer without compromising efficiency.

**BACKGROUND**

**Digital Twin Framework.** In 2017, Dr. Michael Grieves' formal DT framework definition laid the foundation for future advancements (Grieves and Vickers 2017). Initially high-level and theoretical, the concept diverged to suit different field demands. Model choices in these frameworks reflect this variation. For instance, manufacturing sectors, focusing on time efficiency, favor data-driven models for rapid decision-making based on data patterns and trends (Xia et al. 2021). Conversely, sectors like structural health monitoring, requiring high fidelity and interpretability, prefer physics-based models to accurately capture complex behaviors (Ritto and Rochinha 2021). Because DT frameworks are typically conceived with a narrow focus on immediate, specialized needs, most proposed frameworks only satisfy a small set of IRs and FRs necessary for broad industry adoption (Ma et al. 2023). Again, we adopt Gratius et al. 2024's high-level DT framework (Figure 1) to contextualize our proposed work. This framework is distinguished by its broad applicability and adaptability, making it suitable for various applications beyond its initial focus on space habitat systems. Its universal applicability stems from its design, emphasizing dynamic updates and seamless integration between different models.

**Reality Gap.** To address the reality gap, learning methods, due to their ability to adapt and generalize across different conditions, are often used. The exact learning methods are determined by various factors, like application and available data type. For example, reinforcement learning (RL) is particularly prominent in the robotics field when modeling the reality gap, where extensive real-world interactions and high costs associated with real-time training necessitate methods that can handle continuous feedback and adaptation (Salvato et al. 2021). For PMx DTs, one of the most common learning methods to model the reality gap is deep transfer learning (DTL), as it effectively leverages existing knowledge from related domains (e.g., simulation) to improve performance in the target domain (e.g., real-world). Unlike some RL methods that are model-free, DTL is structured around a predefined model architecture. In DTL, measurements or configurations from the simulation and the real world are mainly used in two ways. The first approach uses data from both domains as input to a unified neural network (Xu et al. 2019), or two separate neural networks (Xia et al. 2021), to generate outputs like fault diagnosis, while the second approach utilizes simulation data as inputs to reconstruct the real-world data (Liu et al. 2022). However, the application of DTL to complex assets with multiple components often requires significant domain knowledge for the model structure, making it challenging for intricate systems.

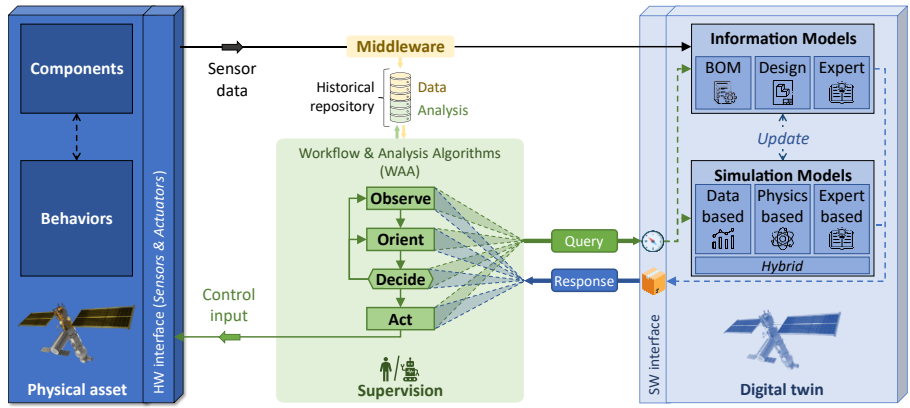

**Figure 1. High-level DT framework proposed by Gratius et al. 2024.**

Despite the approaches implemented, two core limitations exist in the state-of-the-art DT research that tackles reality gaps. The first limitation lies in the quantification of reality gaps. Current literature often models reality gaps as black boxes without distinguishing the contribution to reality gaps from different factors (Xu et al. 2019; Xia et al. 2021). In a DT framework, multiple models interact to replicate the dynamics of the physical asset interacting with the real world.

Inaccurate modeling of interactions with physical entities, such as sensors and environments, increases the discrepancy between the physical and digital domains. Factors like sensor drift ($RG_s$), environmental variability ($RG_e$), and human interactions ($RG_i$) easily cause these inaccuracies, thus contributing to the reality gap in a DT framework. Understanding these components offers targeted strategies to mitigate specific discrepancies and valuable insights into interdependencies between them. The second limitation is the lack of real-to-sim transfer. In any DT framework where simulation and physical realities coexist, focusing solely on sim-to-real transfer neglects the crucial reverse flow of information (Müller et al. 2022). Accurate sim-to-real transfer is vital for transitioning between phases, but to maintain a dynamic and responsive DT during deployment, it is equally important to detach the impact of the reality gap from real-world measurements to accurately reflect and adapt to the evolving or deteriorating nature of the asset.

**DIGITAL TWIN-BASED SIM-TO-REAL AND REAL-TO-SIM TRANSFER**

Existing DT frameworks—including that of Gratius et al. 2024—offer high-level frameworks with minimal exploration of different phases in an asset's lifecycle, meaning the reality gap is typically not explicitly modeled. To overcome this, we expand the Workflow & Analysis Algorithms (WAA) of Gratius et al. 2024's framework by introducing a new module to quantify and mitigate the effects of the reality gap. Data pipelines are designed accordingly between the new module and components within the existing DT framework to ensure DT efficiently utilizes information from different phases.

**Module Development and Integration.** We introduce a data-driven RGA module that correlates real-world sensor data with asset configurations, enabling quantification and management of the reality gap (Figure 2). The historical repository now includes simulation data from the design phase, gathered under various configurations, such as health indices and environmental variables. This data, primarily generated by software models or algorithms like virtual sensors, simulates aspects of physical counterparts in the augmented DT framework to inform design. As the deployment phase starts, the RGA module initiates a sequence of evaluations to ensure efficient calculation of the reality gap. Initially, it queries the DT to check if reality gaps can be calculated using the current DT models (Figure 2, Q1), proceeding only after confirmation (R1). It then checks if design-phase simulation data are in the repository (Q2). If absent (R2 is *False*), the DT is asked to simulate this data (Q3), which is stored in the historical repository (R3). If present (R2 is *True*), the process proceeds. The RGA's data-driven model requests this simulation data (Q4) for pre-training (R4), preparing for effective knowledge transfer during later deployment. In deployment, the RGA predicts asset configurations, requesting corresponding virtual sensor data from simulation models (Q5). The received data (R5) is crucial for identifying reality gaps at sensor locations.

To quantify the reality gap using virtual sensor measurements, we assume that each contributing factor to the reality gap exhibits a distinct, independent normal distribution:

$$RG_s \sim N(\mu_s, \sigma_s^2), \qquad RG_e \sim N(\mu_e, \sigma_e^2), \qquad RG_i \sim N(\mu_i, \sigma_i^2) \tag{1}$$

$$RG_s \perp\!\!\!\perp RG_e \perp\!\!\!\perp RG_i \tag{2}$$

Based on our assumptions, each sensor location on the asset has a unique normal distribution characterizing its reality gap, independent of others. During deployment, the virtual sensor data, reflecting digital domain measurements, are compared with physical sensor data from the real world. Discrepancies between these data sets quantify the reality gap at each sensor location over time. Continual data transfer from physical to digital domains allows the reality gaps to form patterns matching the predefined normal distributions at each location.

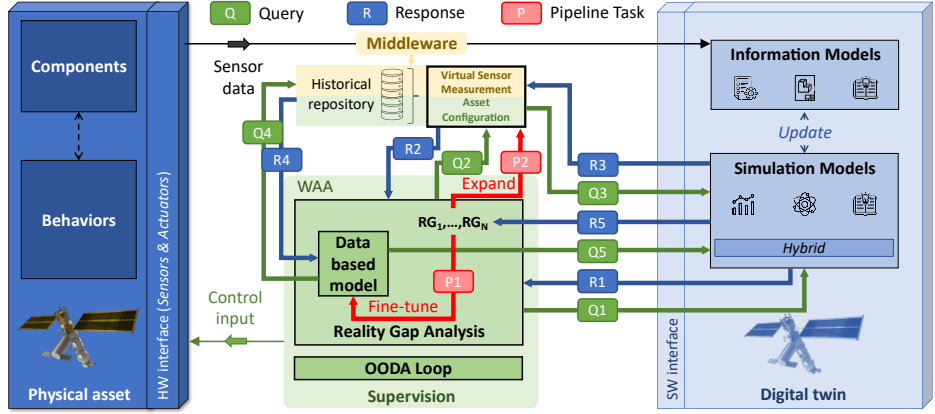

**Figure 2. Augmented DT framework highlighting new module and pipeline functionalities.**

**Data Pipeline.** Upon quantifying the reality gap within the RGA module, the focus turns to minimizing this discrepancy in order to enhance the transferability between the simulation environment and real-world contexts. To achieve this, we implement two primary features within the data pipeline that connect the RGA module, the historical repository, and the DT.

The first feature enables sim-to-real transfer by tailoring the data-driven models in the RGA module to each asset based on its unique reality gaps (Figure 2, P1), by incorporating the quantified reality gap with simulation data from the historical repository to fine-tune the model. This fine-tuning ensures the model accurately represents the projection from an asset's real-world sensor measurements to its configurations. The outcome is a customized RGA module per asset, enabling both accurate sim-to-real transfer and efficient fleet management. The second feature enables real-to-sim transfer by enhancing the historical repository with real-world sensor data, augmenting the existing simulation datasets (Figure 2, P2). To ensure seamless integration of real-world data into a repository of simulation data, we undertake a two-step procedure. Initially, we determine the asset's simulated configuration based on physical sensor readings, then perform a simulation to acquire corresponding virtual sensor data to detach real-world influences from physical sensor data (similar to Figure 2, Q5 and R5). Following this, a confidence-based algorithm assesses the alignment of each sensor's readings with a multivariate normal distribution, derived from the simulation data in the historical repository. Data with significant deviations are added to the repository as new critical conditions. The result is a historical repository with more comprehensive configurations to pre-train RGA modules for future assets.

## CASE STUDY AND DISCUSSION

**Overview.** Next, we provide a case study to illustrate and contextualize the augmentation of the new module and data pipeline within a DT framework. The case study considers the Newell-Simon Bridge (NSB), a pedestrian steel truss bridge on Carnegie Mellon University's campus (Pittsburgh, PA) (Figure 3). This case study considers condition-based monitoring, a subset of PMx that employs sensors installed on the asset to continuously monitor its health status. The aim is to demonstrate our approach's capacity for knowledge transfer between two domains.

For this proof-of-concept study, a simulated environment of the bridge is used, featuring a virtual model with 42 sensors on each outer truss for deformation measurement (highlighted in green in Figure 3). Artificial noise is added to the sensor data to mimic real-world conditions. The simulation is employed in Ansys Mechanical for its structural fidelity and Python for workflow automation, ensuring efficient integration of physical phenomena in the DT framework.

**Performance Metrics.** To assess our approach's effectiveness, we defined metrics for accuracy and efficiency. *Accuracy* is measured using Mean Square Error (MSE) between the real-time data from the physical asset and the simulated data under the same configuration. This metric gauges how closely the DT mirrors the physical asset; a lower MSE indicates better alignment and reduced reality gap. *Efficiency* is evaluated by the model's training time. This metric is straightforward yet powerful, directly relating to the approach's practicality.

**Training and Testing.** For each operational cycle—from training to inference—the data collected in the simulation environment is categorized into three distinct sets: 50% for training, 20% for validation, and 30% for testing, deviating slightly from the standard train-test split by incorporating a validation set that is utilized as a proxy for real-world data entering the digital domain at the start of the deployment phase. It supports our foundational assumption that these gaps follow a normal distribution. Also, this initial set of real-world data allows us to quantify these distributions more accurately. During each cycle, the SGA module is trained using the training set. Subsequently, the validation set is used to ascertain the distribution of the reality gap. The reality gap is then used to further refine and minimize its impact on the model.

**Discussion.** The module and data pipeline introduced are structured into three distinct "Levels of Integration (LoI)," each representing a progressive level of integration and complexity and building upon one another (Figure 3). This structure provides a clear and comparative understanding of the incremental benefits brought by each component of our approach and offers valuable insights into the effectiveness of the augmentation in a real-world engineering context.

While *LoI A* represents an initial implementation of Gratius et al. 2024's framework, it does not fully leverage the complete set of simulation models outlined in their work, failing to account for the reality gap between the physical and digital domains and leading to a lack of high-fidelity synchronization of the data collected from instrumentation onto the DT. As shown in Table 1 of our case study, LoI A mostly succeeds in identifying the general location of critical components, a common outcome observed in our implementation and other state-of-the-art library-based works. Library-based approaches facilitate the transfer of low-level knowledge, such as the approximate fault location, between simulated and real-world scenarios. However, they fall short of accurately conveying more nuanced information, like the severity or precise characteristics of the fault, without a more sophisticated and efficient utilization of data.

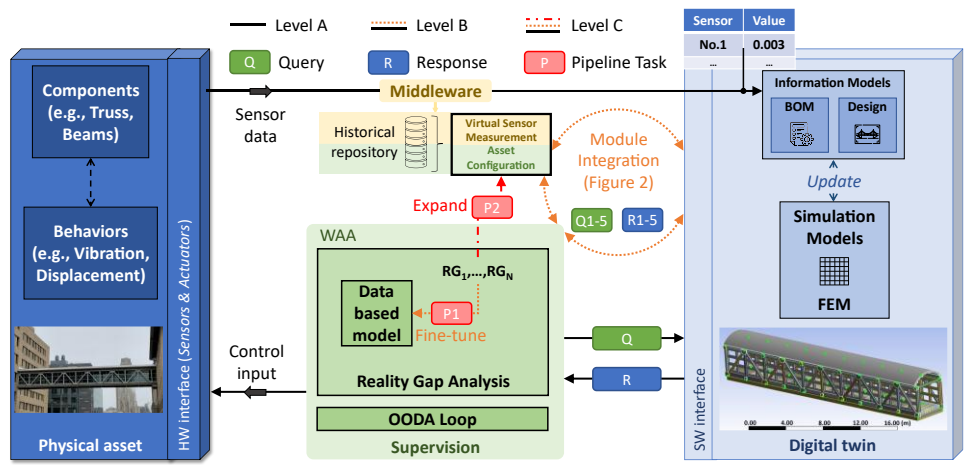

**Figure 3. Step-by-step implementation of the case study.**

**Table 1. Performance compared across different LoIs averaged on 10 random splits. The metrics within each block are represented as (accuracy in MSE, efficiency in second)**

|       | 1 Epoch           | 3 Epochs          | 5 Epochs          | 10 Epochs          |
|-------|-------------------|-------------------|-------------------|--------------------|
| LOI A | (0.000452, 126)   | (0.000432, 377)   | (0.000388, 625)   | (0.000386, 1251)   |
| LOI B | (0.000421, 131)   | (**0.000372,** 401) | (**0.000322,** 670) | (**0.000318,** 1342) |
| LOI C | (**0.000417, 156**) | (0.000377, 471)   | (0.000329, 767)   | (0.000325, 1562)   |

*LoI B* integrates the process of quantifying the reality gap using the validation set. For each sensor, only the reality gap at 95% confidence rate is used to quantify the normal distribution. This process involves combining these distributions with the historical repository to fine-tune the RGA module for sim-to-real transfer. Comparative tests against LoI A reveal that LoI B accurately identifies critical components and reduces the discrepancy between physical and virtual sensor readings (Table 1). Fine-tuning efficiency within LoI B remains relatively unaffected despite the introduction of new data flows.

Building upon LoI B, *LoI C* introduces a new data flow that processes real-world data from the validation set by removing influences from the actual environment. This involves statistically comparing these data with the existing datasets in the historical repository. The comparison is based on the pre-established multi-variant sensor distributions within the repository. Should any sensor data from the validation set—under the corresponding simulation conditions—fall outside the 95% confidence interval of the repository's distribution, it is identified as a novel scenario and subsequently integrated into the repository. This integration recalibrates the sensor distributions in the repository. Table 1 indicates that the performance improvement of LoI C over LoI B is not consistently significant. However, it is important to note that the primary benefit of LoI C extends beyond immediate performance metrics. The expanded repository is essentially real-to-sim transfer, enhancing the accuracy of pre-trained RGA models.

**Example Instance.** To illustrate our methodology, we analyze a single time step measurement from all 42 sensors on the asset (Table 2). This instance-specific visualization contrasts with Table 1's overarching comparison of the different LoIs across the entire dataset. The same real-world configuration underpins LoI A, B, and C, with variations in "Real-world deformation" across groups arising from different injected noise to imitate real-world situations.

**Table 2. Comparison between real-world and simulated deformations of the outside truss.**

| LOI | Real-world deformation | Simulation deformation | Legend |
|---|---|---|---|
| LOI A (MSE: $3.96 \times 10^{-4}$) | 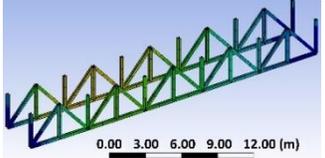 | 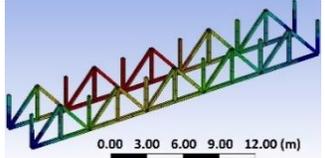 | m<br>0.0120<br>0.0105<br>0.0090<br>0.0075<br>0.0060<br>0.0045<br>0.0030<br>0.0015<br>0 |
| LOI B (MSE: $3.31 \times 10^{-4}$) | 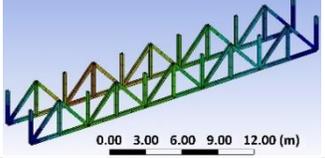 | 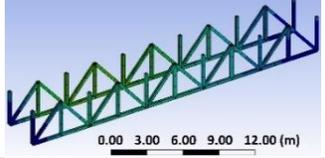 | |
| LOI C (MSE: $3.14 \times 10^{-4}$) | 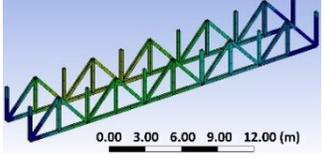 | 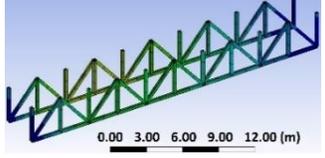 | |

## CONCLUSION

This work investigates the integration of an RGA module within an existing DT framework. The module quantifies the reality gap, enabling the expansion of the original framework to support both sim-to-real and real-to-sim transfer. This effective use of information addresses a limitation of current DT frameworks considering the reality gap, which is the lack of bidirectional knowledge transfer. In addition to providing an overview of the proposed methodology, this paper presents a case study on a condition-based monitoring task. By integrating our approach into the existing DT framework in three different LoIs, we progressively demonstrate the impact of bidirectional knowledge transfer in enhancing monitoring accuracy without compromising efficiency. Looking forward, quantifying the reality gaps originating from different factors is essential, as it provides a more context-based understanding of the reality gap and the interdependencies between these factors. This, in turn, allows for a more targeted and efficient approach to addressing knowledge transfer between simulation and real-world operations.

**ACKNOWLEDGMENTS**

Information developed under this Award was sponsored by the U.S. ARMY Contracting Command under Contract W911NF20D0002 and W911NF22F0014 delivery order No. 4.